\documentstyle[sprocl,psfig]{article}

\def\Journal#1#2#3#4{{#1} {\bf #2}, #3 (#4)}

\def\NPB{{\em Nucl. Phys.} B}
\def\PLB{{\em Phys. Lett.}  B}
\def\PRL{\em Phys. Rev. Lett.}
\def\PRD{{\em Phys. Rev.} D}

\def\be{\begin{equation}}
\def\ee{\end{equation}}
\def\bea{\begin{eqnarray}}
\def\eea{\end{eqnarray}}

\begin{document}

\title{CATALYSIS OF CHIRAL SYMMETRY BREAKING BY EXTERNAL 
MAGNETIC FIELDS IN THREE DIMENSIONAL LATTICE QED}

\author{K.FARAKOS, G.KOUTSOUMBAS} 

\address{Physics Department, National Technical University,\\ 
Zografou Campus, 157 80, Athens, Greece} 

\author{N.MAVROMATOS, A.MOMEN}

\address{Department of Physics, Theoretical Physics, 1 Keble Road,\\
Oxford OX1 3NP, U.K.}

\maketitle\abstracts{The enhancement of the fermionic condensate due to the
presence of both homogeneous and non-homogeneous external magnetic
fields is studied for three-dimensional QED.}

\section{Introduction}

The reason to study the behaviour of the fermionic matter under 
the influence of an external magnetic field \cite{dm,gusynin,leung,wied}
has to do with considerations
connected to the effects of magnetic fields in the early Universe and 
to high $T_c$ superconductivity.

The three-dimensional continuum Lagrangian of the model is given by:
\be
{\cal L} = -\frac{1}{4}(F_{\mu\nu})^2 
+ {\overline \Psi} D_\mu\gamma _\mu \Psi -m {\overline \Psi} \Psi,
\label{contmodel}
\ee
where $D_\mu = \partial _\mu -ig a_\mu^S-i e A_\mu;$ $a_\mu^S$ is a 
fluctuating gauge field, while $A_\mu$ represents the external gauge field.
The main object of interest here is the condensate $<{\overline \Psi} \Psi>,$
which is the coincidence limit of the fermion propagator, $S_F(x,y).$
A first estimate of the enhancement of the condensate arising from the 
external fields may be gained through 
the analysis of the relevant Schwinger-Dyson equation:
\be
S_F^{-1}(p) = \gamma \cdot p  - g \int \frac{d^3k}{ ( 2\pi)^3} 
\gamma^\mu S_F (k)
\Gamma^\nu ( k, p-k) D_{\mu \nu}(p-k)
\label{2.1}
\ee
where $\Gamma^\nu$ is the fermion-photon vertex function and $D_{\mu
\nu}$ is the exact photon propagator.

A result which has been obtained \cite{fm} by approximating the Schwinger-Dyson 
equation for strong magnetic fields is that the dynamically generated 
mass $\Sigma(0)$ is given by:
\be
\Sigma(0) \simeq C ln \left [\frac{\sqrt{e B}}{\alpha}\right ].
\label{dyn}
\ee
There have also been approximations in the regime of smaller  
magnetic fields with interesting results \cite{sphagin}, but for a fully 
quantitative treatment one should rely on the lattice approach \cite{fkm}.

\section{Lattice formulation}

Let us now describe the lattice formulation of the problem. The
lattice action is given by the following formulae:

\bea
&~&S =\frac{\beta_G}{2} \sum_{x,\mu, \nu} F_{\mu \nu}(x) F^{\mu \nu}(x) 
+ \sum_{n,n^\prime} {\overline \Psi}_n Q_{n,n^\prime} \Psi_{n^\prime}
\eea
$$
F_{\mu \nu}(x) \equiv a^S_\mu(x)+a^S_\nu(x+\mu)-a^S_\mu(x+\nu)-a^S_\nu(x)
$$
$$
Q_{n,n^\prime} = \delta_{n,n^\prime}-K \sum_{\hat \mu} 
[\delta_{n^\prime,n+\hat \mu} (r+\gamma_{\hat \mu}) U_{n {\hat \mu}} 
V_{n {\hat \mu}} + \delta_{n^\prime,n-\hat
\mu} (r-\gamma_{\hat \mu}) U_{n-\hat \mu, \hat \mu}^\dagger 
V_{n-\hat \mu, \hat \mu}^\dagger].
$$
The indices $n,~n^\prime$ are triples of integers, 
such as $(n_1,~n_2,~n_3),$
labeling the lattice sites, while $\mu$ denotes directions.
$r$ is the Wilson parameter, $K$ the hopping parameter, 
$U_{n {\hat \mu}} \equiv
e^{i g a \alpha^S_{n {\hat \mu}}},~~V_{n {\hat \mu}} 
\equiv e^{i e a A_{n {\hat \mu}}}$   $\beta_G=\frac{1}{g^2 a}.$ 
$\alpha^S_{n {\hat \mu}}$ represents the statistical gauge
potential and $A_{n {\hat \mu}}$ the
external electromagnetic potential.
$\beta_G$ is related to the statistical gauge coupling constant
in the usual way. On the other hand, we denote by $e$ the 
dimensionless electromagnetic
coupling constant of the external electromagnetic field $U_E(1).$ 
In our treatment we will use na\"ive fermions, so we set $r=0.$

\section{Lattice Results}
We first consider a homogeneous magnetic field and study its effect on
the condensate.
For the construction of a lattice version of a homogeneous  magnetic 
field  we follow \cite{dh}. We will not describe the details
here, but just say that the magnetic field $B$ is given by the expression
$B=m \frac{2 \pi}{N^2},$ for a lattice with spatial extent $N \times N;$
$m$ is an integer and $B$ runs from 0 to $\pi.$ We also note that 
we measured the magnetic field in 
units of its maximal value: thus we used the parameter $b,$ defined 
by: $b \equiv \frac{B}{B_{max}}.$
Since $B_{max} = \pi,$ as explained previously, we get: 
$b = \frac{B}{\pi}$ and $b$ runs from 0 to 1.
We will first present the results for the $T=0$ case.
\begin{figure}[t]
\centerline{\psfig{figure=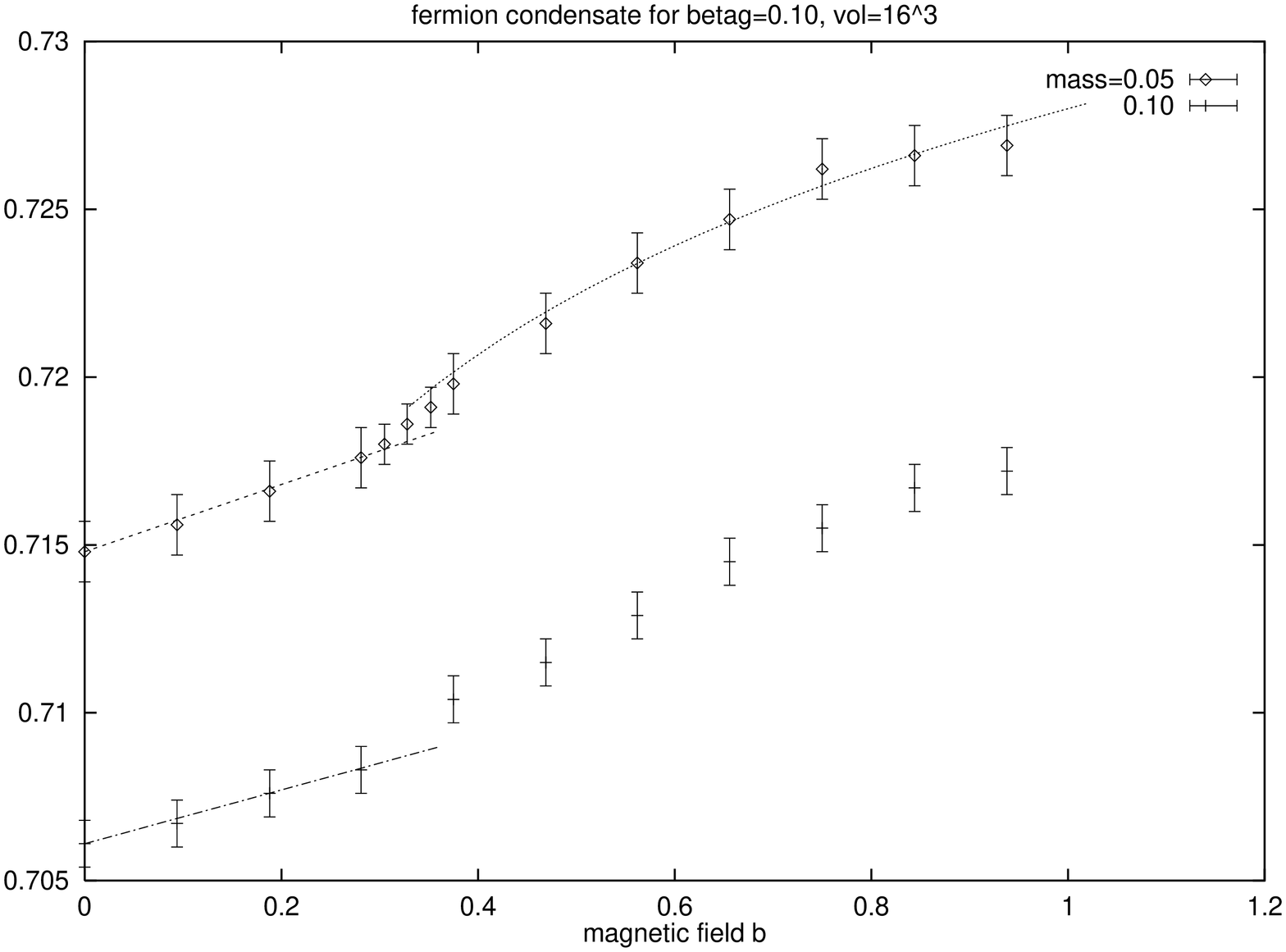,height=4cm,angle=-90}}
\caption{$<{\overline \Psi} \Psi>$ versus the magnetic field
strength at strong coupling for two masses. \label{f1}}
\end{figure}
Figure 1 contains the fermion condensate 
versus the magnetic field for a $16^3$ lattice in the strong coupling 
regime for the statistical gauge field $(\beta_G = 0.10)$ for two
values of the bare mass. For both masses the plot consists of
two parts with qualitatively different behaviour. For $b$ smaller than
about 0.3 we find a linear dependence of the condensate on the external
magnetic field, while for big magnetic fields we find points that
could possibly be fitted to a logarithmic type of curve. The 
logarithmic dependence 
is the one referred to above (equation \ref{dyn})  and has been found 
by an approximate solution of the 
Schwinger-Dyson equations in the regime of strong magnetic fields 
(\cite{fm}). We have included such a logarithmic guide to the eye 
for $m=0.05$ in figure 1.

We now make contact with the results for the model with the statistical 
gauge field turned off. In \cite{fkm} it has been found that 
for big enough $b$ the condensate stopped showing a monotonous
increase with $b,$ at $b=0.5$ it had a local minimum and then had a
succession of maxima and minima, up to $b=1.$ Moreover, there was 
a spectacular volume dependence. One expects, of course that this 
``free" case 
will be reached for big enough $\beta_G.$ 
\begin{figure}[t]
\centerline{\psfig{figure=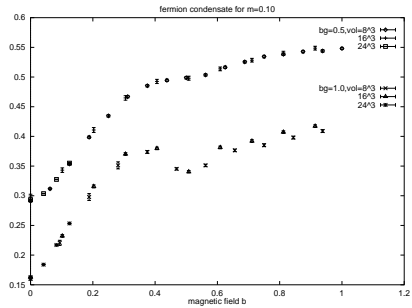,height=4cm,angle=-90}}
\caption{$<{\overline \Psi} \Psi>$ versus the magnetic field
for two big values of the gauge coupling constant 
and three volumes. \label{f2}}
\end{figure}
In figure 2 we show the results 
for $\beta_G=0.5$ and $\beta_G=1.0$ for various volumes. For 
$\beta_G=0.5$ the ``curve" shows the first sign of ``breaking" at 
$b=0.5,$ while at $\beta_G=1.0$ the succession of maxima and minima is
clear. However, there is no detectable volume dependence, so we can 
be sure that, even at this large $\beta_G,$ the limit of switching the
gauge field off has not yet been reached; it will presumably be reached
for even bigger values of $\beta_G.$ 
\begin{figure}[t]
\centerline{\psfig{figure=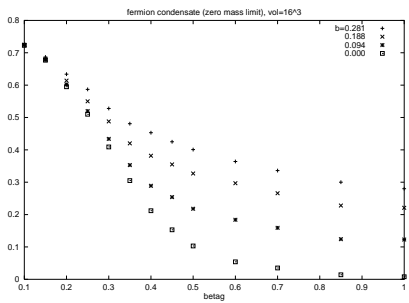,height=4cm,angle=-90}}
\caption{$<{\overline \Psi} \Psi>$ versus $\beta_G$ n the
zero mass limit for four values of the magnetic 
field strength. \label{f3}}
\end{figure}

Figure 3 contains the zero mass limit of the condensate  
versus $\beta_G,$ 
for four values of the external field. We observe that in the 
strong coupling region the b-dependence is rather weak; on the 
contrary, at weak coupling, the condensate is mainly due to
the external field and we find an increasingly big 
b-dependence, as we move to large $\beta_G.$

The last topic will be the study of the response of 3-D QED to a 
non-homogeneous magnetic field; the lattice approach is very efficient here.
The model is considered with the statistical gauge
field turned off. For the construction of the non-homogeneous lattice
magnetic field, we refer the reader to a forthcoming publication (\cite{fkm2}).
\begin{figure}[t]
\centerline{\psfig{figure=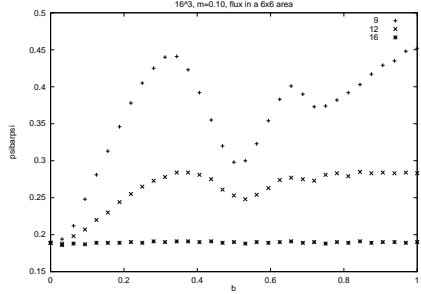,height=4cm,angle=-90}}
\caption{$<{\overline \Psi} \Psi>$ versus magnetic field strength
where the flux is non zero only in a central region extending over
6x6 plaquettes. The condensate at various distances from the center
is shown.  \label{f4}}
\end{figure}

In figure 4 we show the results for a central region of non-vanishing 
flux of extent $6 \times 6.$ More specifically, for the $16^3$ 
lattice we have been using, the region with constant non-zero
flux contains the plaquettes starting at $(n_1,n_2,n_3),$ with 
$6 \le n_1 \le 11$ and $6 \le n_2 \le 11,$ while $n_3$ takes all values.
Note that nothing depends on the value of $n_3.$
The uppermost curve in the figure depicts the result for 
the condensate at the site $(9,9,9).$
We have observed that the results for the sites
$(9,9,9),(9,10,9),(9,11,9),$ which lie totally within the region of 
the non-zero flux, are quite similar. The first substantial change 
takes place at the site $(9,12,9),$ shown in the figure, 
which lies exactly on the boundary
of the above region. The curves corresponding to the sites $(9,13,9),
(9,14,9),~(9,15,9),~(9,16,9)$ dive together to a value 
which is accounted for by the explicit mass term and has very little to do
with the external magnetic field. Only the curve for the last site is shown. 

\section*{Acknowledgements}
K.F. and G.K. would like to acknowledge financial support from the 
TMR project ``Finite temperature phase transitions in particle Physics", 
EU contract number: FMRX-CT97-0122.

\end{document}